# Specific temperature dependence of pseudogap in $YBa_2Cu_3O_{7-\delta}$ nanolayers

A. L. Solovjov[1,2], L. V. Omelchenko[1], V.B. Stepanov[1], R. V. Vovk[3],

H.-U. Habermeier[4,5], H. Lochmajer[6], P. Przyslupski[7] and K. Rogacki[2,6*]

[1] B. I. Verkin Institute for Low Temperature Physics and Engineering of National
Academy of Science of Ukraine, 47 Nauki ave., 61103 Kharkov, Ukraine

[2] International Laboratory of High Magnetic Fields and Low Temperatures, 95 Gajowicka Str., 53-421, Wroclaw, Poland

[3] Physics Department, V. Karazin Kharkiv National University, Svobody Sq. 4, 61077 Kharkiv, Ukraine

[4] Max-Planck-Institut für Festkörperforschung, Heisenbergstrasse 1, D-70569 Stuttgart, Germany

[5] Science Consulting International. Niersteinerstrasse 28, D 70499 Stuttgart, Germany

[6] Institute of Low Temperature and Structure Research,
Polish Academy of Sciences, ul. Okolna 2, 50-422 Wroclaw, Poland

[7] Institute of Physics, Polish Academy of Sciences, al. Lotnikow 32/46, 02-668 Warsaw, Poland

The pseudogap (PG) derived from the analysis of the excess conductivity $\sigma'(T)$ in superlattices and double-layer films of $YBa_2Cu_3O_{7-\delta} - PrBa_2Cu_3O_{7-\delta}$ (YBCO-PrBCO), prepared by pulsed laser deposition, is studied for the first time. The $\sigma'(T)$ analysis has been performed within the local-pair (LP) model based on the assumption of the paired fermion (LPs) formation in the cuprate high-$T_c$ superconductors (cuprates) below the representative temperature $T^* \gg T_c$ resulting in the PG opening. Within the model, the temperature dependencies of the PG, $\Delta^*(T)$, for the samples with different number of the PrBCO layers ($N_{Pr}$) were analyzed in the whole temperature range from $T^*$ down to $T_c$. Near $T_c$, $\sigma'(T)$ was found to be perfectly described by the Aslamasov-Larkin (AL) and Hikami-Larkin (HL) [Maki-Thompson (MT) term] fluctuation theories, suggesting the presence of superconducting fluctuations in a relatively large (up to 15 K) temperature range above $T_c$. All sample parameters were found to change with increase of $N_{Pr}$, finally resulting in the appearance of the pronounced maximum of $\Delta^*(T)$ at high temperatures. The result is most likely due to increasing influence of the intrinsic magnetism of PrBCO ($\mu_{Pr} \approx 4\,\mu_B$) and suggests the possibility to search in that way the change of interplay between the superconductivity and magnetism in cuprates.

## I. INTRODUCTION

Having stimulated a huge amount of publications, the question of what causes superconductivity in the copper-oxide high-$T_c$ superconductors (HTSCs) is widely considered to be one of the great challenges of the condensed-matter physics [1–4]. Gradually, it has become clear that usual electron-phonon interaction, proposed by the BCS theory [5], is hardly possible to account for formation of the superconducting (SC) Cooper pairs at such high temperatures [6] and any additional interaction mechanism is to be taken into consideration [4, 7–10].

Both cuptrates [2–4, 9] and FeAs-based superconductors [Fe-pnictides (FePns)] [11, 12] are known to be magnetic materials in their parent state. In this state, strong on-site repulsion in cuprates prevents electron motion and turns the material into a Mott insulator with a long-range antiferromagnetic (AF) order at low doping [8, 13]. As the hole concentration (doping) increases, the long-range AF order rapidly destroys and superconductivity emerges [7–9, 13]. However, the corresponding theories [14, 15] as well as neutron measurements [16, 17] show that, after the long-range AF order breaks down, the short-range AF correlations persist up to the rather high doping level. In FePns the AF order of spin-density-wave-type (SDW-type) is found to coexist with superconductivity in a wide range of doping [9–12]. This leaves little doubt that pairing and scattering in these materials are both affected by the low-energy AF fluctuations [8, 18–20]. The facts suggest that magnetic correlations are the most probable additional interaction mechanism for the Cooper pair formation in both kinds of HTSCs, but the question still remains controversial.

Importantly, apart from the high $T_c$s, cuprates possess the so-called pseudogap (PG) which opens below any representative temperature $T^* \gg T_c$ ([2–4, 8–10, 21] and references therein). Various models have been put forward to explain both the pairing mechanism and PG phenomenon in HTSCs [2, 3, 8–10, 21–23], including various forms of electron pairing [3, 22–25], spin-fluctuations [26–28], interplay with charge fluctuations [4, 29] and even spin-charge separation scenarios [30, 31]. We believe the PG to be due to the formation of preformed pairs [local pairs (LPs)] [2, 3, 22, 23, 32]. Nevertheless, the pairing mechanism being responsible for the electron coupling at very high temperatures very likely can be of a magnetic type [1, 8, 10, 30, 33]. But the issue as for the PG physics still remains controversial, suggesting the study of the interplay between superconductivity and magnetism to be one of the challenging problems of high-temperature superconductivity [34].

To clarify this issue, in this paper we study the fluctuation conductivity (FLC) and PG in $YBa_2Cu_3O_{7-\delta}$-$PrBa_2Cu_3O_{7-\delta}$ (YBCO-PrBCO) superlattices (SLs) and YBCO-PrBCO double-layer films [(so-called "sand-wiches" (SDs)] with different layer composition, prepared by the pulsed lased deposition [35] and sequential high-pressure sputtering [36]. $Pr^{+3}$ atoms are known to have an intrinsic magnetic moment, $\mu_{eff} \approx 3.58\mu_B$ [37] and $\mu_{eff} \approx 2\mu_B$ as the PrBCO compound [38]. That is why such compounds are considered to be very promising in studying the change of interplay between superconductivity and magnetism in HTSCs, which has to increase along with increase of $N_{Pr}$. We expected to reveal the influence of this intrinsic magnetism on FLC and the PG, especially seeing that no direct evidence of this influence on the PG has been reported so far.

## II. EXPERIMENT

The progress in thin film preparation technology [35, 39, 40] makes high-quality superlattice thin films available for detailed analysis. The YBCO-PrBCO SLs have been grown on SrTiO₃ (100) substrates by pulsed laser ablation, as described elsewhere [35]. X-ray and Raman-scattering analysis have shown that all samples are excellent films with the c-axis perfectly oriented perpendicular to the $CuO_2$ planes. Next the SLs were lithographically patterned and chemically etched into well-defined $1.68 \times 0.2$ mm² Hall-bar structures. SLs with layer periodicity $\Lambda$ equal to 4 YBCO × 1 PrBCO (4Y×1Pr), 7Y×7Pr and 7Y×14Pr (samples SL1, SL2, SL3) have been analyzed. The total number of $\Lambda$ is 20 for all SLs. But only the YBCO layers were taken into account in calculating $\rho(T)$. Importantly, the thickness of one layer is assumed to be $d = 11.7$ Å $= c$, which is the c-axis lattice parameter [41]. For more information on the properties and quality of the superlattices studied, see Refs. [7, 42, 43].

To show the more universal character of PGs, two SDs (heterostructures) with composition 40PrBCO × 50YBCO (sample SD1) and 40Pr-BCO × 20YBCO (sample SD2), where the numbers imply the width in nanometers of the PrBCO and YBCO, respectively, have also been studied. The PrBCO/YBCO heterostructures were deposited using a sequential high-pressure sputtering from stoichiometric targets [39]. The SDs were deposited at $T = 770\,^{\circ}C$ in 3 mbar oxygen pressure at $(LaAlO_3)_{0.3}(Sr_2TaAlO_6)_{0.7}$ substrates. The thickness of deposited films was controlled by the deposition time of respective targets. The single YBCO layers with the thickness of 50, 20 nm are marked Y50 and Y20, respectively, whereas bilayers with 40 nm Pr layer and 50, 20 nm YBCO layer are marked Pr40Y50 (sample SD1) and Pr40Y20 (sample SD2), respectively. To perform contacts golden wires glued to the structure pads with silver epoxy. Contact resistance below 1 $\Omega$ was obtained. A fully computerized setup utilizing the four-point probe technique was used to measure the in-plane resistivity $\rho_{ab}(T) = \rho(T)$.

More information on the properties and quality of the heterostructures studied can be found in Refs. [44–46].

It will be observed that the crystal cell of PrBCO is isostructural to YBCO but PrBCO is an insulator (see Ref. [47] and references therein). Importantly, in preparing $Y_{1-x}Pr_xBa_2Cu_3O_{7-\delta}$ (YPrBCO) films PrBCO is evaporated simultaneously with YBCO. As a result, one gets an YBCO matrix including a set of randomly distributed insulating PrBCO cells which produce multiple structural defects. As $x$ increases, the resistivity rapidly increases, too, while $T_c$ and the charge carrier density $n_f$ decrease. Eventually, YPrBCO becomes an insulator when $x \geq 0.7$ [47]. This occurs as a result of a Fehrenbacher-Rice (FR) energy band formation in which the free charge carriers in the $CuO_2$ planes have to localize [48]. This process has little influence on the $CuO$ chains, the number of holes in which remains practically constant [49]. As a consequence, the concentration of oxygen in YPrBCO also remains constant. Thus, the exploration of YPrBCO compounds permits the study of the HTSCs property variation immediately upon a change of $n_f$.

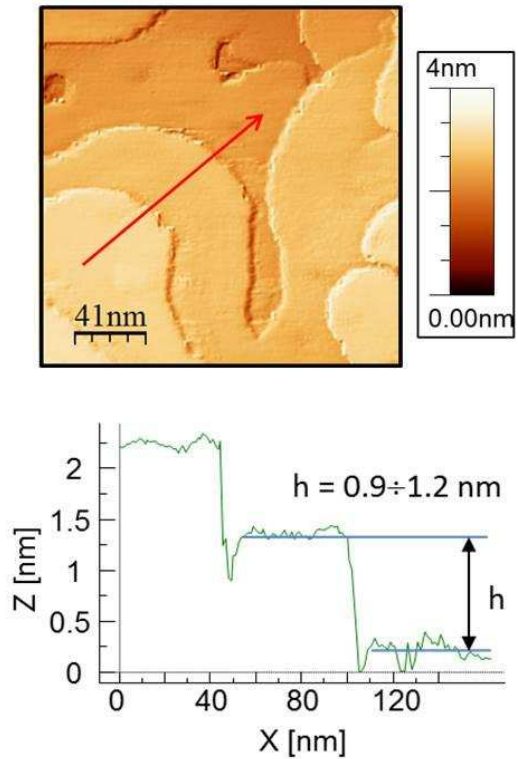

FIG. 1: (top panel) STM image of the L90Y50 bilayer showing a step-like structure of the surface YBCO layer; the scanned area is $207 \times 207\,nm^2$. (bottom panel) STM linear profile taken along the arrow; average height of the observed steps is from 0.9 to 1.2 nm, which is close to the c-axis lattice parameter of the YBCO compound.

In the process of the YBCO-PrBCO SLs preparation after several layers of PrBCO ($N_{Pr}$) are deposited, the corresponding number of layers of the optimally doped (OD) YBCO ($N_Y$) is evaporating and so on [35]. As a result, one gets a set of the well-structured superconducting YBCO nanolayers embedded into the insulating PrBCO matrix [35, 47] whose properties can be studied. In our case the maximum thickness of the nanolayer is $d_0 = 11.7 \times 7 \approx 82$ Å. In contrast to the YPrBCO films, this time, except for a proximity effect, the PrBCO has no direct influence on the YBCO layers [48]. The SDs, being constructed from two relatively thick YBCO and PrBCO layers, behave in the similar way.

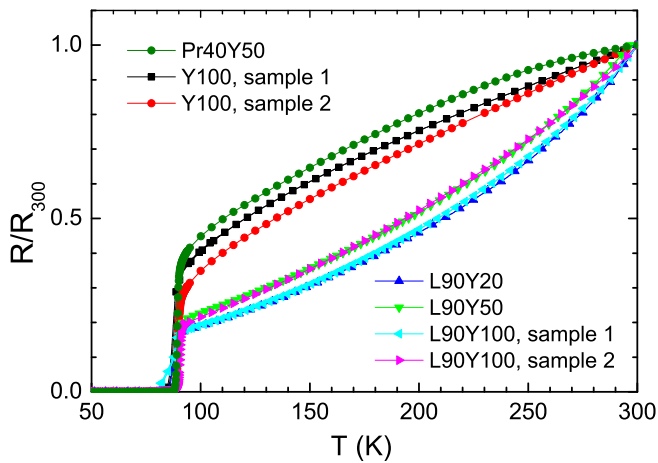

FIG. 2: Normalized resistance versus temperature for the Pr40Y50 bilayer, together with two Y100 monolayers and four L90Yd bilayers with d = 20, 50, and 100, for comparison. Numbers in the samples names indicate the thickness of the individual layers in nm. $R_{300}$ indicates the resistance at 300 K.

For better characterization of our samples, the scanning tunneling microscope (STM) studies have been performed for several samples. Figure 1 shows an example image of the surface of an L90Y50 bilayer, which is the 50-nm-thick YBCO layer deposited on the 90 nm (La,Sr)MnO3 (LSMO) layer. The observed step-like structure is a consequence of the epitaxial growth of our sample. The image reveals that the maximum change of sample thickness is less than 4 nm. Average height of the observed steps is about 1.1 nm, which is close to the c-axis lattice parameter of the YBCO compound (c = 1.168 nm). This picture is typical for larger crystallites, independently which part of the sample ($5 \times 10 \, mm^2$) is studied. Uniformity of smaller crystallites is better. The STM surface studies of the bilayers show a topology similar to that observed for highquality single-layer YBCO films.

Resistance as a function of temperature of several monolayers and heterostructures is shown in Fig. 2. Charac-teristic $R(T)/R_{300}$ dependencies are observed for different kinds of samples, where $R_{300}$ is the resistance at 300 K, indicating that the sputtering process is well controlled and the sample properties are reproducible. These layers show a relatively sharp transition to the superconducting state, $\Delta T < 3K$ (criterion $0.1R_n - 0.9R_n$, where $R_n$ is the resistance at the normal state), except for L90Y100 (sample 1) which shows $\Delta T \sim 8K$. For all layers, the zero resistance was observed within the accuracy of the measurement error. Heterostructures characterized by the corresponding R(T) dependence and sharp transition to the superconducting state ($\Delta T < 1.5K$) have been taken for further investigations.

Figure 3 shows temperature dependencies of resistivity both for SLs and SDs. With the increase in the relative number of $N_{Pr}$ against $N_Y$ within the layer stack $\Lambda$, $T_c$ of the samples gradually decreases. The process becomes more visible when ratio $(N_{Pr})/(N_Y) = N^* > 2$. In this case, the significant reduction of $T_c$ is followed by nonmetallic resistivity temperature dependence [47]. Obviously, to be analyzed in terms of the LP model [9, 50–54] only samples with metallic $\rho(T)$, like the aforesaid SLs and SDs, have been studied.

Somewhat surprisingly no logical $T_c$ vs $N^*$ dependence is revealed either for the SLs nor for SDs. It is assumed to be due to $N^* \le 2$ in both cases. Nevertheless, the increase of $N_{Pr}$ deeply affects the shape of the resistivity curves (Fig. 3). Really, both SL1 ($N^* = 0.25$) and SD1 ($N^* = 0.8$) demonstrate $\rho(T)$ being close to that usually observed for unadulterated YBCO films [55, 56]. Above the pseudogap temperature $T^*$, i.e. in the normal state, it is characterized by pronounced linear $\rho(T)$, which ranges at least up to $\sim 340$ K, as was measured for the SDs. But with increase of $N^*$, the resistivity curves are found to demonstrate excessive resistivity being more pronounced in the case of SL3 and SD2 ($N^* = 2$). However, no noticeable buckling of the resistivity curves, being typical for FePns and slightly doped cuprates, is observed below $T^*$, and $\rho(T)$ as before remains linear above $T^*$ (Fig. 3). Unexpectedly, the values of $T_c$ remain rather high and even increase a little bit in the case of both SL3 and SD2, whereas $T^*$s are found to increase noticeably (Fig. 3 and Table I). This is in contrast to $Y_{1-x}Pr_xBa_2Cu_3O_{7-\delta}$ film with $x = 0.1$ (sample L100 [47]) in which the FR localization mechanism has to work. In this case the deep reduction of $T_c$ down to 78 K is followed by the very pronounced increase of resistivity and considerable decrease of $T^*$ down to 178 K. As the FR mechanism does not work in the SLs and SDs it is rather tempting to ascribe the revealed peculiarities of the resistivity (Fig. 3) to the essential magnetism of PrBCO. The results of the FLC and PG analysis were expected to confirm the conclusion.

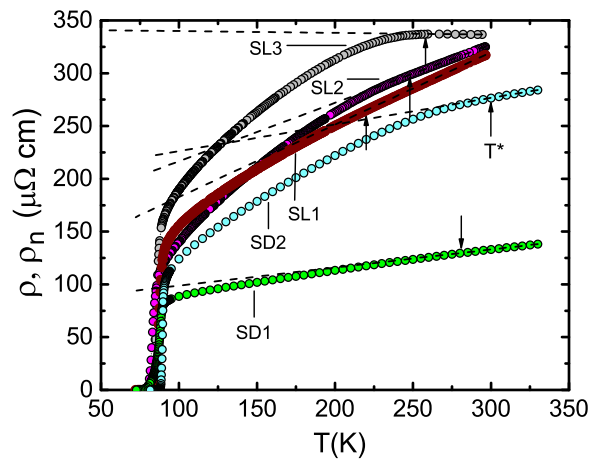

FIG. 3: In-plane resistivity $\rho$ of $YBa_2Cu_3O_{7-\delta}$-$PrBa_2Cu_3O_{7-\delta}$ SL's and sandwiches as a function of T. Straight dashed lines designate extrapolated $\rho_N(T)$. Arrows determine $T^*$ for all samples.

## III. RESULTS AND DISCUSSION
### A. Fluctuation conductivity

As appears from Fig. 3, at $T \leq T^*$, resistivity of all samples downturns from its linear dependence observed at higher temperatures. This leads to the emergence of pronounced excess conductivity

$$\sigma'(T) = \sigma(T) - \sigma_N(T) = [1/\rho(T)] - [1/\rho_N(T)], \quad (1)$$

as a difference between measured $\rho(T)$ and linear normal-state resistivity $\rho_N(T)$ extrapolated to the low-$T$ region [26, 57, 58]. As usual, $T^*$ is taken at the point where the experimental resistivity curve starts to downturn from the high-temperature linear behavior [9, 57–60]. The more precise approach to determine $T^*$ with accuracy $\pm 0.5$ K is to explore the criterion $\rho_N(T) = aT + \rho_0$ [52], where $a$ is the slope of the extrapolated $\rho_N(T)$ and $\rho_0$ is its intercept with the y axis. Apparently, above $T^*$, where $\rho = \rho_N$, $[\rho(T) - \rho_0]/aT = 1$ and its deviation from unity just determines $T^*$. This approach is illustrated in Fig. 4, which shows sample SD2 as an example. Both methods give the same $T^*$.

As was convincingly shown by NMR [61] and angle resolved photoemission spectroscopy (ARPES) [62] experiments, in cuprates at $T \leq T^*$ not only resistivity decreases but electronic density of states (DOS) at the Fermi level also starts to decrease. The state with a partially reduced DOS above $T_c$ is just a pseudogap [4, 9, 10]. Thus, one may conclude that, if there were no decrease of DOS below $T^*$ resulting in the PG opening, resistivity $\rho(T)$ would remain linear down to $T_c$, as is observed in conventional superconductors [61]. Hence, the excess conductivity $\sigma'(T)$, which appears as a result of the PG opening, has to contain information about the pseudogap. To get the information, a special approach based on the local-pair (LP) model has been developed [9, 54]. In accordance with the model, the PG is believed to appear due to the LPs formation at $T \leq T^*$, accordingly regarded as a pseudogap temperature. At high temperatures $T \lesssim T^*$ the LPs are believed to appear in the form of the so-called strongly bound bosons (SBBs) which obey the theory of Bose-Einstein condensation (BEC) [3, 50–53]. Note that either of the coupling mechanisms proposed by different aforementioned models can be responsible for the pairing at such high temperatures. But the proper mechanism of LP pairing still remains unknown. At $T_{pair} << T^*$ SBBs transform into fluctuating Cooper pairs (FCPs) [22, 23, 32, 58], thus demonstrating the BEC-BCS transition [9, 54] predicted by the theory [50–53]. This is a specific property of the HTSCs, which is the consequence of the extremely short coherence length in cuprates [2, 3, 9, 10, 50].

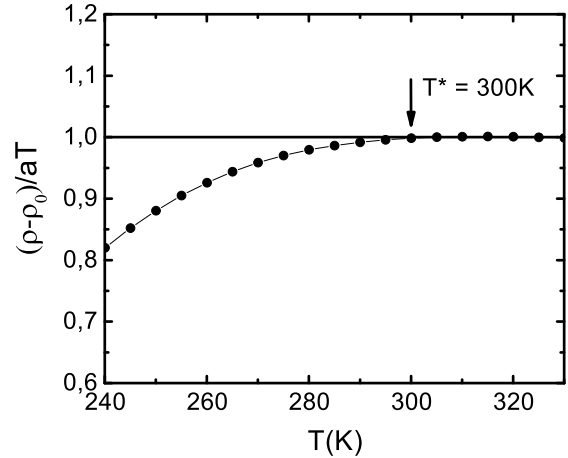

FIG. 4: $[\rho(T) - \rho_0]/aT$ as a function of temperature for sample SD2 (dots), which determines $T^* = 300 \pm 0.5$ K. The straight line is to guide the eye.

$\sigma'(T)$, found with a help of Eq. (1) for all five aforementioned samples, has been analyzed within the LP model [9, 54] paying more attention to the possible difference of the revealed results in comparison with those obtained for YPrBCO (sample L100) [47] and YBCO (sample F1) [63] films regarded as reference objects. Importantly, all studied samples have been treated in the identical way. However, to somehow simplify our discussion we will consider the analysis performed for SL3 ($7Y \times 14Pr$) as a reference sample, and finally compare results for all samples studied (see Table I and II). To initiate analysis we have to find the mean-field critical temperature $T_c^{mf}$, which determines the reduced temperature [64]

$$\varepsilon = (T - T_c^{mf}) / T_c^{mf}. \quad (2)$$

Here, $T_c^{mf} > T_c$ is the critical temperature in the mean-field approximation, which separates the FLC region from the region of critical fluctuations or fluctuations of the SC order parameter $\Delta$ directly near $T_c$ (where $\Delta < k_B T$), neglected in the Ginzburg-Landau (GL) theory [5]. As was convincingly shown (see Ref. [9, 54, 65, 66] and references therein), near $T_c$, $\sigma'(T)$ is always extrapolated by the standard equation of the Aslamasov-Larkin (AL) theory [67] with the critical exponent $\lambda = -1/2$ which determines FLC in any three-dimensional (3D) system,

$$\sigma'_{AL3D} = C_{3D} \frac{e^2}{32\,\hbar\,\xi_c(0)} \varepsilon^{-1/2},\qquad(3)$$

where $C_{3D}$ is a numerical factor used to fit the data to the theory [9, 59] and $\xi_c(T)$ is a coherence length along the c axis [64]. This is because, in cuprates near $T_c$ $\xi_c(T)$ becomes larger than $d$ [64], and FCPs acquire an ability to interact in the whole sample volume, thus forming the 3D state. This means, in turn, that the conventional 3D FLC is always realized in HTSCs as $T \to T_c$ [64, 68]. From Eq. (3), one can easily obtain $\sigma'^{-2} \sim (T - T_c^{mf})/T_c^{mf}$. Evidently, $\sigma'^{-2} = 0$ for $T = T_c^{mf}$ [59]. Moreover, when $T_c^{mf}$ is properly chosen, the data in the 3D fluctuation region near $T_c$ are always fit by Eq. (3) [63].

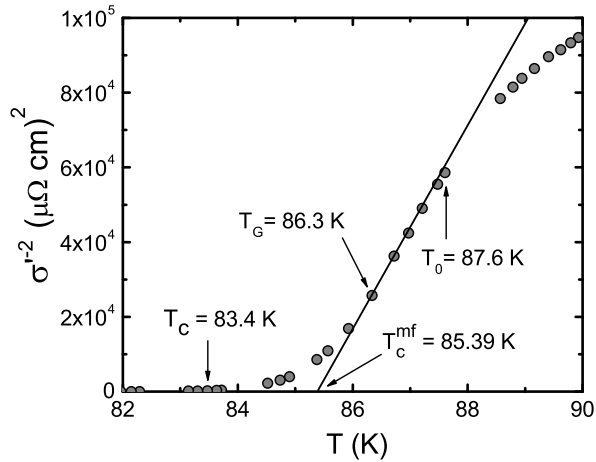

FIG. 5: Temperature dependence of $\sigma'^{-2}(T)$ for YBCO-PrBCO superlattice SL3 (dots), which determines $T_c^{mf} = 85.39\ K$. The straight line is to guide the eye.

Fig. 5 displays the $\sigma'^{-2}$ vs $T$ plot for SL3 (dots). The intercept of the extrapolated linear $\sigma'^{-2}$ with $T$-axis determines $T_c^{mf} = 85.39 \pm 0.01\ K$. Also shown is Ginsburg temperature $T_G = 86.3 \pm 0.02\ K$ down to which $\sigma'(T)$ obeys the fluctuation theories. Above the three-dimensional to two-dimensional (3D-2D) crossover temperature $T_0 = 88.6 \pm 0.02\ K$, the data deviates right from the line suggesting the presence of 2D Maki-Thompson (MT) fluctuation contribution into the $\sigma'(T)$ [64]. At the crossover temperature $T_0 \sim \varepsilon_0$ ($ln\varepsilon_0$ in Figs. 4 and 5) the coherence length $\xi_c(T) = \xi_c(0)\varepsilon^{-1/2}$ has to amount to $d$ [63, 64, 68]. This yields

$$\xi_c(0) = d\sqrt{\varepsilon_0}\qquad(4)$$

and permits the determination of $\xi_c(0)$ which is one of the important parameters of the LP model analysis.

Excess conductivity $\sigma'$, measured for all studied samples, is plotted in Figs. 6 and 7 as a function of $\varepsilon$ in customary double logarithmic scale. The result for sample SL3 is designated as dots in Fig. 4. As expected, above $T_G$ ($ln\,\varepsilon_G \approx -4.8$) and up to $T_0$ ($ln\,\varepsilon_0 \approx -3.3$) $ln\sigma'$ vs $ln\epsilon$ is fit by the 3D AL fluctuation term (3) (Fig. 6, solid line 1) with $\xi_c(0) = 2.24 \pm 0.02$ Å (Table I) determined by Eq. (4), $C_{3D} = 1.94$, and $d = c \approx 11.7$ Å [41], as mentioned above. The better is the sample structure quality, the closer $C_{3D}$ is to 1 [59, 63]. The variation of the $C_{3D}$ values, observed in Table II, is likely due to ambiguity in the width of the conducting layers in the SLs. Found $\xi_c(0) = 2.24 \pm 0.02$ Å is about 1.36 times that obtained for the YBCO film F1 ($T_c = 87.4$ K) [63] but 1.5 times less than $\xi_c(0)$ found for the YPrBCO film L100 ($T_c = 87.4$ K) [47], pointing out the expected difference between SLs and the reference films (Table I and II).

Because $\xi_c(T) = \xi_c(0)\varepsilon^{-1/2}$ has to decrease with increase of temperature, the 3D state is lost at $T > T_0$, where $\xi_c(T) < d$ [68]. But in the range $T_0 < T < T_{01}$ it is still larger than the distance between the internal conducting $CuO_2$ planes $d_{01} \sim 4$ Å [41]. As a result, $\xi_c(T)$ still connects the internal planes by the Josephson interaction, and the system is believed to be in a quasi-2D state [63, 64, 68]. That is why, above $T_0$ and up to $T_{01} \approx 97$ K (designated as $ln\varepsilon_{01} = -2$ in Fig. 6) $ln\sigma'$ vs $ln\varepsilon$ is perfectly fit by the 2D MT fluctuation term of the HL theory [64]

$$\sigma'_{MT} = \frac{e^2}{8\,d\,\hbar} \frac{1}{1-\alpha/\delta} ln\left((\delta/\alpha)\frac{1+\alpha+\sqrt{1+2\,\alpha}}{1+\delta+\sqrt{1+2\,\delta}}\right)\varepsilon^{-1},\qquad(5)$$

which dominates in the 2D fluctuation region $T_0 < T < T_{01}$ [9, 64, 68] (Fig. 6, dashed curves 2). Accordingly,

TABLE I:
The sample parameters.

| Sample | $\rho(100K)$ $\mu\Omega cm$ | $T_c$ (K) | $T_c^{mf}$ (K) | $T_{01}$ (K) | $T_G$ (K) | $\Delta T_{fl}$ (K) | $\xi_c(0)$ Å |
|---|---|---|---|---|---|---|---|
| F1 | 148.0 | 87.4 | 88.46 | 97.3 | 88.1 | 9.2 | $1.65 \pm 0.02$ |
| SL1 | 155.2 | 85.1 | 87.7 | 95.3 | 88.3 | 7.0 | $2.13 \pm 0.02$ |
| SL2 | 139.8 | 80.8 | 82.74 | 95.6 | 83.8 | 11.8 | $2.87 \pm 0.02$ |
| SL3 | 189.4 | 83.4 | 85.39 | 96.7 | 85.9 | 10.8 | $2.24 \pm 0.02$ |
| SD1 | 88.6 | 85 | 88.2 | 100 | 88.6 | 11.4 | $1.86 \pm 0.02$ |
| SD2 | 124 | 88.5 | 89.22 | 105 | 89.4 | 15.6 | $1.25 \pm 0.02$ |
| L100 | 226.0 | 78.0 | 82.1 | 92.7 | 82.6 | 10.1 | $3.35 \pm 0.02$ |

above $T_{01}$, where $\xi_c(T) < d_{01}$, the pairs are confined within the CuO$_2$ planes and there is no interplane interaction now [68]. This is why, above $T_{01}$, the fluctuation theories do not describe the experiment, as is clearly seen in Figs. 6 and 7. In this way, it follows that $\xi_c(T_{01}) = d_{01}$. Thus, $T_{01}$ determines the range of the SC fluctuations $\Delta T_{fl} = T_{01} - T_G$ [22, 32] which is about 11 K above $T_c$ is the case of SL3.

In Eq. (5)

$$\alpha = 2\left[\frac{\xi_c(0)}{d}\right]^2 \varepsilon^{-1} \qquad (6)$$

is a coupling parameter,

$$\delta = \beta\frac{16}{\pi\hbar}\left[\frac{\xi_c(0)}{d}\right]^2 k_B T \tau_\phi \qquad (7)$$

is the pair-breaking parameter, $\tau_\phi$ is defined by [63]

$$\tau_\phi \beta T = \pi\hbar/8k_B\varepsilon = A/\varepsilon \qquad (8)$$

is the phase relaxation time, and $A = 2.988 \cdot 10^{-12}$ sK. The factor $\beta = 1.203(l/\xi_{ab})$, where $l$ is the mean-free path and $\xi_{ab}$ is the coherence length in the $ab$ plane, considering the clean limit approach ($l > \xi$) being typical for cuprates [9, 63].

It should be noted that the MT fluctuation description does not work in the case of YPrBCO film (sample L100 in the Tables). In contrast to the SLs, up to $T_{01} \approx 92.7$ K ($ln\varepsilon_{01} \approx -2.05$) the data are well described by the Lawrence-Doniach term of the HL theory

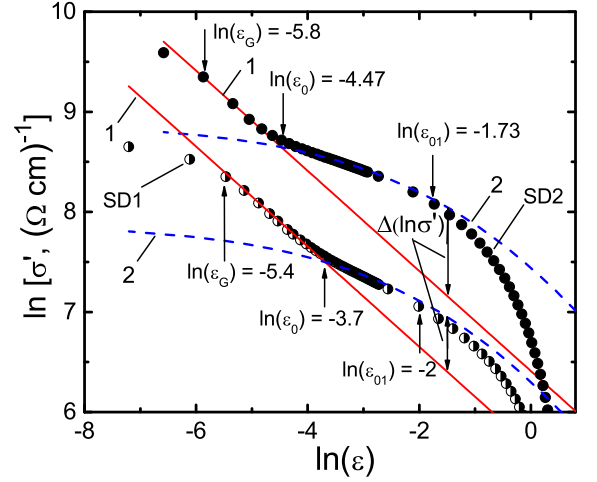

FIG. 7: $ln\sigma'$ vs $ln\varepsilon$ for YBCO-PrBCO sandwiches SD1 (half-filled circles) and SD2 (dots) compared with fluctuation theories: the label 1 identifies 3D AL (solid lines) and the label 2 identifies 2D MT (dashed curves).

[64] now, which is typical for YBCO films with defects [69]. This finding confirms the above conclusion that, in YPrBCO films (so-called "alloys"), PrBCO produces multiple defects in the YBCO matrix resulting in noticeable increase of $\rho(T)$, $\xi_c(0)$ and corresponding decrease of $T_c$, $T^*$ and $\Delta^*(T_c)$ (see the tables).

Also shown in Fig. 6 are the data for SL1 (squares) and SL2 (half-filled circles). In the case of SL1 one layer of PrBCO is believed to produce defects in the YBCO matrix. It results in a deeply suppressed MT fluctuation contribution, which is typical for the YBCO films with defects [69]. With increasing $N_{Pr}$ the MT fluctuation contribution becomes much more pronounced. Finally, in the case of SL3 the distance $\Delta(ln\sigma')$ between the data and the AL term extrapolated down to the 2D fluctuation region becomes noticeably large. The effect of $\Delta(ln\sigma')$ enlargement is much more pronounced in the case of sandwiches (Fig. 7). For the first time, the noticeable increase of $\Delta(ln\sigma')$ was observed in Fe-pnictide SmFeAsO$_{0.85}$ ($T_c = 55$ K) [9]. Thus, the effect is believed to be the first evidence of the enhanced influence of magnetism in YBCO-PrBCO compounds.

Nevertheless, for all samples, the reliable values of $\xi_c(0)$ (Table I), which are determined by the corresponding crossover temperature $T_0$ (Eq. (4)), can be obtained from the fits. Note, that the usual fitting approach with $d = 11.7$ Å, $\alpha$ determined by Eq. (6) and $\varepsilon = \varepsilon_0$ in Eq. (8), which we used to analyse pure YBCO films [63], still works perfectly in the case of SL1 and SL2, where the magnetic influence is relatively small (Fig. 4). However, to perform the theoretical 2D MT fit for SL3 and both sandwiches, we have to use $\varepsilon = \varepsilon_{01}$ in Eq. (8) now. Accordingly, Eq. (6) has to be rewritten as

$$\alpha = 2\left[\frac{\varepsilon_{01}}{\varepsilon}\right] \qquad (9)$$

taking equality $\xi_c(0) = d\sqrt{\varepsilon_0} = d_{01}\sqrt{\varepsilon_{01}} = 2.24$ Å into account and assuming $d = d_{01}$. In the case of sandwiches the more reasonable values of $d_{01} = \xi_c(0)(\sqrt{\varepsilon_{01}})^{-1}$ and the largest $\Delta T_{fl}$ were obtained (Table I). It is in contrast with the SLs, for which somewhat enhanced $d_{01}$ by comparison with these reported by structural studies [41], were found (Table II).

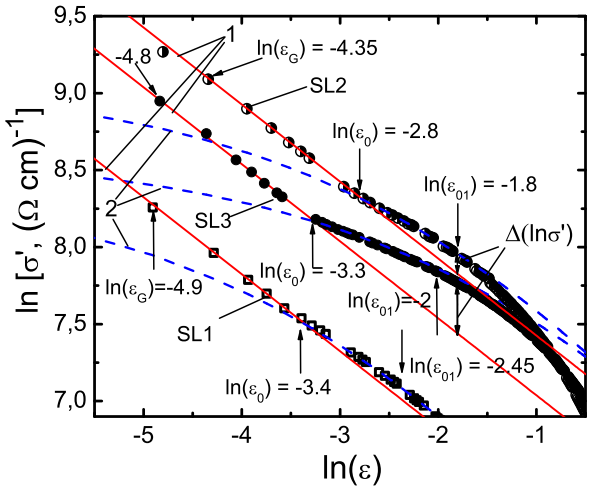

FIG. 6: $ln\sigma'$ vs $ln\varepsilon$ for YBCO-PrBCO superlattices SL1 (squares), SL2 (half-filled circles) and SL3 (dots) compared with fluctuation theories: the label 1 identifies 3D AL (solid lines) and the label 2 identifies 2D MT (dashed curves).

Thus, just $\varepsilon_{01} \propto T_{01}$ governs Eq. 5 in the case of enhanced magnetic interaction. For every samples the value of $ln\varepsilon_{01}$ is distinctly seen on the plot, thus allowing the possibility of amount $T_{01}$, which determines the range of the SC fluctuations above $T_c$, as mentioned above. In accordance with the theory [2, 22], $T_{01}$ is the temperature up to which the order-parameter phase stiffness, as well as the superfluid density $n_s$, have to maintain in HTSCs, as confirmed by experiment [32, 70]. This means that in the temperature range from $T_c^{mf}$ up to $T_{01}$, FCPs behave in a good many way like superconducting, but incoherent pairs (short-range phase correlations [2, 3, 8, 22]). This results in specific behavior of the cuprates, which is unconventional from the viewpoint of "classical" superconductivity [32, 70–73].

### B. Pseudogap analysis

Evidently, to get any information about PG from the excess conductivity one needs an equation which specifies a whole experimental curve from $T^*$ down to $T_c^{mf}$ and contains the parameter $\Delta^*$ in explicit form. In cuprates, $\Delta^*$ is referred to as a pseudogap parameter which is mostly due to the local pair formation, as mentioned above. Thus, $\Delta^*(T)$ has to reflect the peculiarities of the LPs interaction along with the decrease of temperature from $T^*$ down to $T_c$ [9, 74, 75]. In YBCO-PrBCO compound, $\Delta^*$ is assumed to be due to both local pair formation and magnetic interaction. Thus, its temperature dependence is expected to somehow reflect the complex interplay between superconducting fluctuations and magnetism which is of primary importance to comprehend the principles of the coupling mechanism in HTSCs.

Because of absence of a complete fundamental theory, we have applied our LP model approach to the PG analysis. The equation for $\sigma'(\varepsilon)$ has been proposed in Ref. [74] with respect to the local pairs

$$\sigma'(\varepsilon) = \frac{e^2 A_4 \left(1 - \frac{T}{T^*}\right) \left(exp\left(-\frac{\Delta^*}{T}\right)\right)}{(16 \hbar \xi_c(0) \sqrt{2 \varepsilon_{c0}^*} \sinh(2\varepsilon / \varepsilon_{c0}^*))}. \qquad (10)$$

Here, the dynamics of both pair-creation $(1 - T/T^*)$ and pair-breaking $exp(-\Delta^*/T)$ below $T^*$ has been taken into account in order to correctly describe the experiment [9, 66, 74]. Solving Eq. (10) regarding $\Delta^*(T)$ one can readily obtain

$$\Delta^*(T) = T \, ln \frac{e^2 A_4 \left(1 - \frac{T}{T^*}\right)}{\sigma'(T) \, 16 \, \hbar \, \xi_c(0) \, \sqrt{2 \, \varepsilon_{c0}^* \, \sinh(2\varepsilon / \varepsilon_{c0}^*)}}, \qquad (11)$$

where $A_4$ is a numerical factor which has the meaning of the C-factor in FLC theory [9, 21, 59, 65, 74] and $\sigma'(T)$ is the experimentally measured excess conductivity over the whole temperature range from $T^*$ down to $T_c^{mf}$. As can be seen from Fig. 8, Eq. (10) (dashed curve 1) fits the data very well, thus demonstrating the validity of the LP model approach. The same result was obtained for

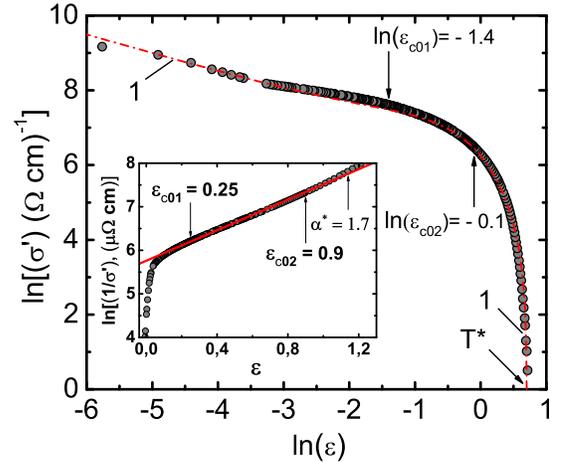

FIG. 8: $ln\sigma'$ vs $ln\varepsilon$ (dots) for SL3 plotted in the whole temperature range from $T^*$ down to $T_c^{mf}$. The dashed curve (1) fits the data with Eq. (10). Insert: $ln\sigma'^{-1}$ as a function of $\varepsilon$. Solid line indicates the linear part of the curve between $\varepsilon_{c01} \simeq 0.25$ and $\varepsilon_{c02} \simeq 0.9$. Corresponding $ln\varepsilon_{c01} \simeq -1.4$ and $ln\varepsilon_{c02} \simeq -0.1$ are marked by the arrows at the main panel. The slope $\alpha^* = 1.7$ determines the parameter $\varepsilon_{c0}^* = 1/\alpha^* = 0.59$.

### TABLE II:
The sample parameters.

| Sample | $C_{3D}$ | $d_{01}$ Å | $\Delta (ln\sigma')$ | $T^*$ (K) | $\Delta^*(T_c)_{exp}$ (K) | $\Delta^*(T_c)_{th}$ (K) | $D^*$ |
|--------|----------|-----------|---------------------|-----------|---------------------------|--------------------------|-------|
| F1 | 1.0 | 5.2 | 0.12 | 203 | 219 | 218 | $5 \pm 0.1$ |
| SL1 | 0.95 | 7.4 | 0.07 | 220 | 215.6 | 212.7 | $5 \pm 0.1$ |
| SL2 | 3.85 | 7.8 | 0.12 | 248 | 202.3 | 202 | $5 \pm 0.1$ |
| SL3 | 1.94 | 6.3 | 0.4 | 258 | 201.9 | 208.5 | $5 \pm 0.1$ |
| SD1 | 0.7 | 5.1 | 0.49 | 281 | 211.6 | 212.5 | $5 \pm 0.1$ |
| SD2 | 1.0 | 3.0 | 0.82 | 300 | 221.5 | 221.2 | $5 \pm 0.1$ |
| L100 | 0.83 | - | - | 172 | 198 | 195 | $5 \pm 0.1$ |

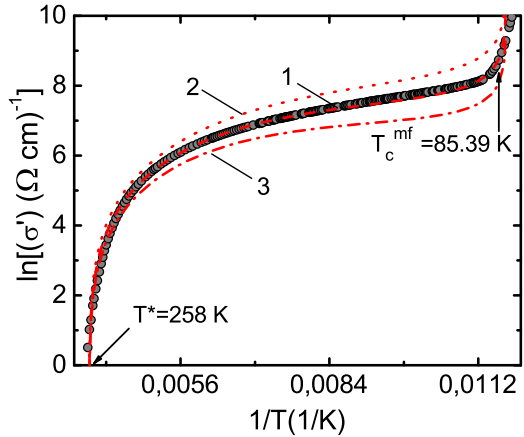

FIG. 9: $ln\sigma'$ vs $1/T$ (dots) for SL3 plotted in the whole temperature range from $T^*$ down to $T_c^{mf}$. The dashed curves are fits to the data with Eq. (10). The best fit is obtained when Eq. (11) is calculated with $\Delta^*(T_c)$=209 K ($D^* = 2\Delta^*(T_c^{mf})/k_B T_c = 5.0$ (curve 1). Curves 2 and 3 correspond to $D^*$=3.6 and 6.4 , respectively.

all other samples studied. From our point of view the result has to confirm the conclusion by means of Eq. (11), $\Delta^*(T)$ has to properly reflects the properties of the pseudogap [9, 74].

Apart from $T^*$, $T_c^{mf}$ and $\xi_c(0)$ determined above, both Eq. (10 and 11) contain several additional parameters important for the analysis. These are the theoretical parameter $\varepsilon_{c0}^*$ [76], the numerical factor $A_4$ and $\Delta^*(T_c)$, which is the PG value at $T_c^{mf}$. Nevertheless, all the parameters can be determined directly from the experiment within our approach. First, in the range of $ln\varepsilon_{c01} < ln\varepsilon < ln\varepsilon_{c02}$ (Fig. 8) or accordingly $\varepsilon_{c01} < \varepsilon < \varepsilon_{c02}$ (106.5 K < T < 162 K for SL3, see insert in Fig. 8), $\sigma'^{-1} \sim \exp(\varepsilon)$ [76]. This exponential dependence turned out to be the common feature of the HTSCs [9, 74, 76, 77]. As a result $\ln(\sigma'^{-1})$ is a linear function of $\varepsilon$ with a slope $\alpha^*$=1.7, which determines parameter $\varepsilon_{c0}^* = 1/\alpha^*$=0.59 (insert in Fig. 8).

To find $A_4$ we calculate $ln\sigma'(ln\varepsilon)$ using Eq. (10) across the whole temperature interval up to $T^*$ and fit it to experiment in the range of 3D AL fluctuations near $T_c$ (Fig. 8, dashed curve 1), where $ln\sigma'(ln\varepsilon)$ is a linear function of the reduced temperature $\varepsilon$ with a slope $\lambda = -1/2$ [9, 74]. As can be seen in the figure, the fit with $A_4 = 35$ is very good, suggesting the perfect structural quality of the studied YBCO nanolayers.

Next, in our consideration $\Delta^*(T_c) = \Delta(0)$ is assumed, where $\Delta(0)$ is the SC gap at $T = 0$ [72, 73]. Thus the equality $D^* = 2\Delta^*(T_c)/k_B T_c =2\Delta(0)/k_B T_c$ is to occur. Finally, to estimate $\Delta^*(T_c)$, which we use in Eq. (10) to determine $A_4$, we plot $ln\sigma'$ as a function of $1/T$ (Fig. 9, dots) [74, 75]. In this case the slope of the theoretical curve [Eq. (10)] turns out to be very sensitive to the value of $\Delta^*(T_c)$ [9, 74]. Despite the influence of magnetism, the fit is completely good—again most likely due to the perfect YBCO layers structure. The best fit is obtained when $2\Delta^*(T_c)/k_B T_c = 5.0 \pm 0.1$ (Fig. 9, dashed curve 1), which is believed to be close to the strong-coupling limit usually observed for cuprates [9, 78–80]. The result suggests that $\Delta^*(T_c)/k_B \approx 209$ K ($\approx 18$ meV). It seems to be reasonable seeing that measured $T_c = 83.4$ K is somewhat low. Thus, all parameters needed to calculate $\Delta^*(T)$ are determined now. Just the same approach was used to determine the corresponding parameters for all other studied samples (Table I and II). Figure 10 (gray dots) displays $\Delta^*(T)$ calculated for SL3 by using Eq. (11) with the following set of parameters derived from experiment: $T_c^{mf} = 85.39$ K, $T^* = 258$ K, $\xi_c(0) = 2.24$ Å, $\varepsilon_{c0}^* = 0.59$, and $A_4 = 35$ (Table I). Also shown in the figure are the $\Delta^*(T)$ dependencies for all other studied samples calculated by using the corresponding sets of found parameters.

As can be seen from the figure, $\Delta^*(T)$ obtained for SL1 demonstrates a wide maximum at $T_{max} \approx 138$ K with $\Delta_{max}^* \approx 250$ K. The shape of the curve is rather close to that found for unadulterated YBCO films [55, 63]. With increase of the Pr content, $\Delta_{max}^*$ decreases whereas $T^*$ increases. Simultaneously, the pronounced maximum of $\Delta^*$ appears at high temperatures, which gradually increases along with $N_{Pr}$. The sandwiches demonstrate just the same behavior (Fig. 10).

For the first time such $\Delta^*(T)$ dependence with a descending linear region was observed for $SmFeAsO_{0.85}$ be-tween $T_s = 150$ K and $T_{SDW} = 130$ K and is believed to be the more noticeable feature of the magnetic influence in the HTSCs [9, 74]. Thus, one may conclude that the specific $\Delta^*(T)$, with pronounced maximum at high $T$, can be attributed to the enhanced magnetic interaction in the YBCO-PrBCO compounds. The enhancement of the magnetic interaction can also explain the observed increase of the $T^*$s, if assumed, as mentioned above, the pairing mechanism at high temperatures to be mostly of the magnetic type. It is worth noting that the shape of $\Delta^*(T)$ for both SL3 and SD2 is actually the same over the whole temperature range down to $T_c^{mf}$ suggesting the same mechanism of the interplay between superconductivity and magnetism. Thus we may conclude that, despite the strong influence of magnetism, our LP model approach has allowed us to obtain rather reasonable and self-consistent results.

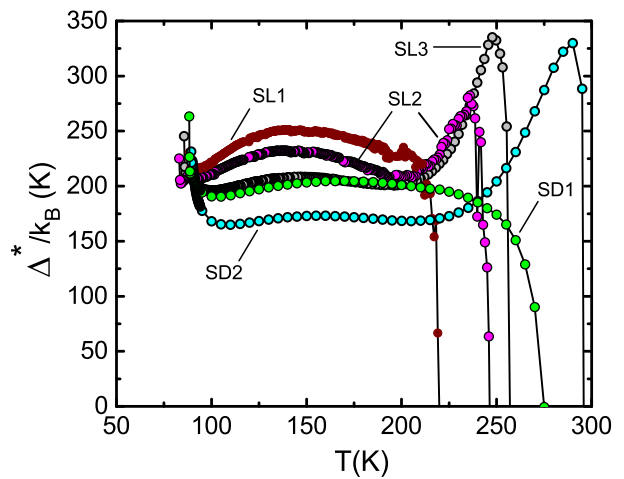

FIG. 10: Temperature dependencies of pseudogap $\Delta^*$ for all samples studied, analyzed with Eq. (11). The maximum at high temperatures gradually increases along with increase of $N_{Pr}$ (see the text).

To be more sure we have compared results (Fig. 11) with those obtained for SmFeAsO$_{0.85}$ (Fig. 11, red dots) and EuFeAsO$_{0.85}$F$_{0.15}$ (Fig. 11, blue dots). The results of the comparison are plotted in the figure in double-reduced scale. It turned out that both the range of the descending linear region and its slope are the same for all shown samples. In SmFeAsO and EuFeAsO, as well as in whole other pictides, the representative temperature $T_s$ corresponds to the structural transition, whereas $T_{SDW}$ corresponds to the antiferromagnetic (AF) ordering of a spin density wave (SDW). In the case of SmFeAsO$_{0.85}$ [9] the linear drop of $\Delta^*(T)$ was qualitatively explained within the Machida-Nokura-Matsubara (MNM) theory developed for the superconductors in which the AF ordering may coexist with the superconductivity, such as, for example, RMo$_6$S$_8$ (R = Gd, Tb, and Dy) [81]. In accordance with the MNM theory in such compounds $\Delta(T)$ linearly drops below $T_N < T_c$ due to the formation of an energy gap of SDW on the Fermi surface which partially suppresses the SC gap. Because the AF gap saturates at lower temperatures, $\Delta(T)$ gradually recovers its value upon increasing the SC condensation energy. The similar $\Delta^*(T)$ behavior in SmFeAsO$_{0.85}$ but above $T_c$ (Fig. 11) was considered to be an additional evidence for the LPs existence in the FeAs-based superconductors [9, 74]. Really, it was assumed that, in accordance with the MNM theory, the order parameter of the local pairs $\Delta^*$ is suppressed below $T_s$ by the low-energy magnetic fluctuations [19, 82–84] resulting in the observed linear drop of $\Delta^*(T)$ followed by the transition to the SDW state. Similarity of the results suggests the likely presence of the magnetic fluctuations in YBCO-PrBCO compounds, too, which are believed to be responsible for the $\Delta^*(T)$ shape at high temperatures (Fig. 11).

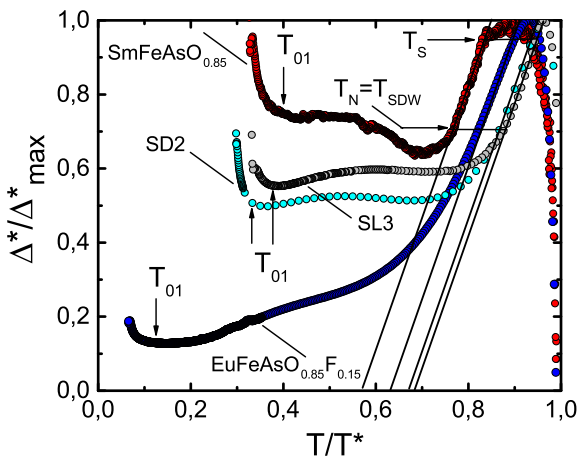

FIG. 11: $\Delta^*(T)/\Delta_{max}$ as a function of $T/T^*$ for studied YBCO-PrBCO superlattice SL3 and sandwich SD2 compared with reference Fe-pnictide samples SmFeAsO$_{0.85}$ ($T_c \approx 55$ K) [9] and EuFeAsO$_{0.85}$F$_{0.15}$ ($T_c \approx 11$ K) [54]. Solid lines with equal slope correspond to the linear $\Delta^*(T)$ region for all samples. Horizontal lines designate its length which lasts between the structural transition temperature $T_s$ and the SDW ordering temperature $T_{SDW}$ (see the text). Arrows at $T_{01}$ designate the ranges of SC fluctuations.

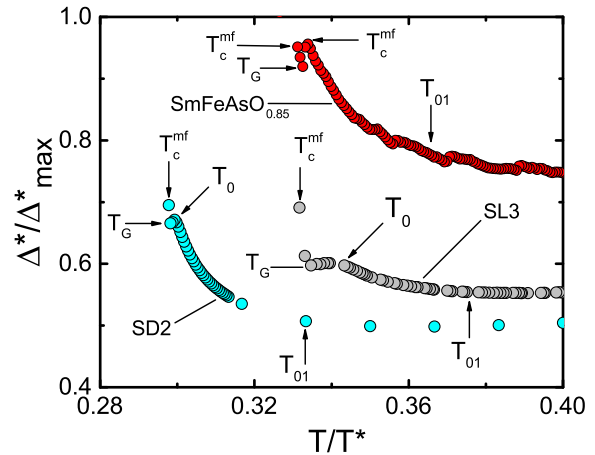

FIG. 12: $\Delta^*(T)/\Delta_{max}^*$ as a function of $T/T^*$ for studied YBCO–PrBCO superlattice SL3 and sandwich SD2 compared with reference Fe-pnictide sample $SmFeAsO_{0.85}$ ($T_c \approx 55 K$) [9]. All representative temperatures are designated by the arrows (see the text). Arrows at $T_{01}$ designate the ranges of SC fluctuations.

Importantly, below $T_{01}$, that is, in the region of the SC fluctuations, all samples also demonstrate similar $\Delta^*(T)$ behavior (Fig. 11). Really, in all samples $\Delta^*(T)$ starts to noticeably increase below $T_{01}$, as detailed in Fig. 12, which is actually a zooming of the corresponding part of Fig. 11. As can be seen in Fig. 12, in all samples including EuFeAsO$_{0.85}$F$_{0.15}$, whose data are out of the range shown, $\Delta^*(T)$ rapidly increases below $T_{01}$ demonstrating maximum at about $T_0$. Then $\Delta^*(T)$ unexpectedly decreases down to $T_G$ which limits the range of the mean-field theory validity [5]. Below $T_G$ it again sharply increases, because the superconductor transfers into the range of the critical fluctuations ($\Delta < k_B T$) below $T_c^{mf}$. We would like to emphasize that just the same $\Delta^*(T)$ dependence also was observed in pure YBCO films [74] and FeSe polycrystals [85]. Thus, one may conclude that all high-$T_c$ superconductors behave in the same way at the approach to the $T_c$ from above. There is always the noticeable range of SC fluctuations just above $T_c$ [22, 32, 71] in which LPs behave like incoherent Cooper pairs, and excess conductivity is described by the AL [67] and HL [64] fluctuation theories.

Summarizing the results, it is rather tempting to conclude that the basic mechanism of the interplay between the superconductivity and magnetism looks suspiciously the same in all compounds where superconductivity can coexist with magnetism. It is very likely that the possibility of the SDW state formation even in the YBCO compounds under some special conditions has to be taken into account. Recently, similar ideas as for the SDW state in YBCO, but at the low doping level, as well as the possibility of the Fermi surface reconstruction below $T^*$, were discussed in Ref. [4]

## IV. CONCLUSION

The $YBa_2Cu_3O_{7-\delta}$-$PrBa_2Cu_3O_{7-\delta}$ (YBCO-PrBCO) superlattices (SLs) and YBCO-PrBCO double-layer films ("sandwiches" or SDs) with different layer composition have appeared to be very promising model objects in studying the interplay between superconductivity and magnetism in HTSCs. The interplay is believed to increase along with increase of the $N_{Pr}$ because PrBCO has intrinsic magnetic moment $\mu(PrBCO) = 1.9 \pm 0.1 \mu_B$ [9]. Importantly, in the case of the SLs the very thin ($7 \times 11.7$ Å $\approx 82$ Å) but homogeneous YBCO nanolayers embedded into insulating PrBCO matrix can be studied.

Independently of the layer composition, near $T_c$, the excess conductivity $\sigma'(T)$ derived from the resistivity measurements was shown to be perfectly described by the 3D AL term and 2D MT term of the conventional fluctuation theories [64, 67] (Figs. 6 and 7). Thus, there is a range of SC fluctuations near $T_c$, which lasts up to the representative temperature $T_{01} \approx 15$ K above $T_c$ in good agreement with theory [22]. In accordance with the theory, in this temperature range the stiffness of the order parameter wave function of the superconducting Cooper pairs has to be maintained. As a result, in a definite temperature interval up to $T_{01}$, the LPs behave like the SC Cooper pairs, which is a specific unusual property of HTSCs [22, 32, 70, 71].

With the increase of the ratio $N^* = (N_{Pr})/(N_Y)$, $T_c$ somewhat decreases, whereas $\rho(T)$ and $T^*$ noticeably increase (Fig. 3). The coherence length $\xi_c(0)$ also was found to decrease, suggesting the likely decrease of the ab-plane coherence length $\xi_{ab}$ which determines the size of the LPs. It has to result in the noticeable increase of the bonding energy $\varepsilon_b \propto 1/\xi_{ab}^2$ [50–53], and finally in the observed increase of $T^*$ (Figs. 3, 10). Simultaneously, the noticeable enhancement of the 2D fluctuation conductivity $\Delta(ln\sigma')$ was observed, pointing out the expected increase of the magnetic interaction in SL3 (7Y-14Pr, $N^* = 2$; Fig. 6) and SD2 (20Y-40Pr, $N^* = 2$; Fig. 7).

For the first time, the analysis of the pseudogap in such objects has been performed within the LP model based on the assumption of the LPs formation in cuprates below $T^*$. In both SL1 ($N^* = 0.25 < 1$) and SD1 ($N^* = 0.8 < 1$) the temperature dependence of PG, $\Delta^*(T)$, resembles $\Delta^*(T)$ usually observed for unadulterated YBCO films [9, 74]. However, with the increase of $N_{Pr}$ (SL3 and SD2) the shape of the $\Delta^*(T)$ curve changes and becomes close to that observed for Fe pnictides, with a sharp $\Delta^*(T)$ maximum at high temperatures followed by the linear descending region (Fig. 10). In SLs every YBCO nanolayer ($w \approx 80$ Å) undergoes a noticeable magnetic influence from the two nearest PrBCO layers with a width $14 \times 11.7$ Å $\approx 164$ Å (SL3). Thus, the specific $\Delta^*(T)$ dependence revealed at high temperatures is considered to be a consequence of the enhanced magnetic interaction. Nevertheless, below $T_{01}$; that is, in the region of the SC fluctuations, all samples, regardless of magnetic interaction, demonstrate the very similar $\Delta^*(T)$ behavior detailed in (Fig. 12). Thus, all high-$T_c$ superconductors are found to behave in the same way at they approach $T_c$ from above.

The comparison with results of the PG analysis obtained for $SmFeAsO_{0.85}$ and for $EuFeAsO_{0.85}F_{0.15}$ (Fig. 11) has shown that the range of the descending linear region and its slope are also the same for both YBCO-PrBCO compounds and FePns, suggesting a similar mechanism of magnetic interaction in different kinds of HTSCs. In FePns the representative temperature $T_s$ corresponds to the structural transition and $T_{SDW}$ corresponds to the AF ordering of spin density wave (SDW) type [11]. Thus, one may conclude that it is very likely that, in YBCO-PrBCO compounds with enhanced Pr content, as well as in FePns, the transition into the SDW state with decrease of temperature seems to be very possible. Recently, the possibility of both SDW state and the charge density wave (CDW) state in unadulterated YBCO was widely discussed in Ref. [4]. Finally, it is very tempting to conclude that the basic mechanism of the interplay between the superconductivity and magnetism could be the same in different kinds of magnetic superconductors.

## ACKNOWLEDGEMENTS

The authors would like to thank Dr. G. Urbanik for STM sample characterization. The work of K. Rogacki and P. Przysłupski was supported by the Wroclaw Research Centre EIT+ within the project NanoMat (POIG.01.01.02-02-002/08) co-financed by the European Regional Development Fund (Operational Programme Innovative Economy, 1.1.2).

---

* Electronic address: k.rogacki@int.pan.wroc.pl